\documentclass[preprint,showpacs,12pt]{revtex4-1} 

\usepackage{epsfig}

\begin{document}

\title{Dynamics of fronts in optical media
with linear gain and nonlinear losses}

\author{Izzat M. Allayarov}
\author{Eduard N. Tsoy$^*$}

\affiliation{Physical-Technical Institute of the Uzbek
Academy of Sciences,\\
Bodomzor yuli st. 2-B, Tashkent, 100084, Uzbekistan\\
$^*$Corresponding author: etsoy@uzsci.net}

\begin{abstract}
  The dynamics of fronts, or kinks, in dispersive media with gain and
losses is considered. It is shown that the front parameters, such as
the velocity and width, depend on initial conditions. This result is
not typical for dissipative systems. For exponentially decreasing
initial conditions, the relations for the front parameters are found. A
presence of the global bifurcation, when a soliton solution is replaced
by the front solution, is demonstrated. It is also shown that in order
to observe fronts, the front velocity should be larger than the
characteristic velocity of the modulational instability.
\end{abstract}

\maketitle

\section{Introduction}
\label{sec:i}

  In conservative media, light can propagate as a stable non-diverging
beam due to a balance between diffraction (dispersion) and
nonlinearity. Such a localized wave with a bell-shaped intensity
profile in the transverse direction corresponds to a soliton, see
e.g.~\cite{Akh97,Kiv03}. Solitons are investigated actively as
important objects of nonlinear optics.

  Another type of localized waves, namely, fronts (shock waves, kinks,
transition waves)  are studied less in optics
\cite{Agr92,Dar98,Car00,Fre03}. A front is a local variation of the
field that connects two different uniform states
\cite{Saa92,Saa03,Ara02,Vas87,Vol00,Akh05}. Often, one state is stable,
while the other is unstable (metastable). Then, the stable state
expands into the unstable one, and the front corresponds to a
transition region between the two states.

  The dynamics of fronts in dissipative media is a subject of extensive
studies in physics~\cite{Saa92,Saa03,Ara02,Vas87,Vol00,Akh05}. An
important model, that describe front propagation, is the nonlinear
diffusion (ND) equation~\cite{Saa92,Vas87,Vol00}:
\begin{equation}
  {\partial_z u} - D \partial_{x}^2 u +f(u) = 0,
\label{nde}
\end{equation}
where $u(x,z)$ is the real-valued field (e.g. pressure or
concentration), $x$ and $z$ are the spatial and evolutional (time)
coordinates, respectively, $D$ is the diffusion coefficient, and $f(u)$
is a nonlinear function. A well-known example of the ND equation is the
Fisher-Kolmogorov equation~\cite{Fis37,Kol37}, where $f(u)= u(u-1)$.
The study of this equation, started in the first half of the twentieth
century, has revealed the main properties of fronts in {\em diffusive}
media. Further development of the theory was connected, in particular,
with the study of complex-valued fields and the combined effect of
diffusion and {\em dispersion}. The corresponding model can be written
as
\begin{equation}
  {i \partial_z \psi} + (\beta_r + i \beta_i) \partial_x^2 \psi
  + F(\psi, \partial_x \psi, \partial_x^2 \psi \dots) = 0,
\label{evol}
\end{equation}
where $\psi(x,z)$ is a complex field, $F$ is a function of $\psi$ and
it's derivatives, and $\beta_r$ ($\beta_i$) characterizes dispersion
(diffusion). A well studied example of model~(\ref{evol}) is the
complex Ginzburg-Landau equation that describes pattern formation in a
variety of physical systems~\cite{Akh97,Saa92,Ara02,Akh05}.

  Typically, model~(\ref{evol}), as well as~(\ref{nde}), possesses a
family of front solutions with different spatial distributions and
velocities. However, many particular examples of the model demonstrate
the following property. A wide class of initial conditions evolve into
a well-defined front with a specific velocity $v^{\star}$. This
velocity depends only on the parameters of the model. In other words,
the system ``selects'' a particular front from a variety of all
possible fronts.

  There are different approaches for calculation of the parameters of
the selected front. The most developed approach is the theory of
marginal stability (MS) that states the following~\cite{Saa92,Saa03}.
Let an evolutional equation in form~(\ref{evol}) has a front solution.
Then the selected front velocity $v^{\star}$ is found from the
dispersion relation of the linearized Eq.~(\ref{evol}), $\omega=
\omega(k) \equiv \omega_{\mathrm{r}}(k)+ i\, \omega_{\mathrm{i}}(k)$.
Namely~\cite{Saa92,Saa03},
\begin{equation}
  v^{\star}=  \omega_{\mathrm{i}} (k^{\star}) /
    k_{\mathrm{i}}^{\star},
\label{vstar}
\end{equation}
where the complex wavenumber $k^{\star}\equiv k_{\mathrm{r}}^{\star} +
i\, k_{\mathrm{i}}^{\star}$ is obtained from
\begin{equation}
  \left. {d \omega(k) \over d k} \right|_{k= k^{\star}} =
  \omega_{\mathrm{i}} (k^{\star}) / k_{\mathrm{i}}^{\star}.
\label{keq}
\end{equation}
The result~(\ref{vstar}) is not proven rigorously, but it works for
many types of evolutional equations.

  In this paper, we analyze a limit of purely dispersive media without
diffusion. This situation is typical in optics. We consider a beam
propagation in media with linear gain and nonlinear losses (two-photon
absorption). Such parameters are characteristic for laser systems. We
demonstrate that at some conditions, the beam expansion in such media
is related to the propagation of two fronts moving in opposite
directions. Then, the beam evolution can be restored, to some extent,
from of the front parameters. Unfortunately, the theory of MS is not
useful for analysis of fronts in media with pure dispersion. We find
that the front parameters in dispersive media depend not only on the
the system parameters, but also  on {\em initial conditions}, see
Sec.~\ref{sec:sw}.

  The optical beam dynamics in nonlinear media is described by the
generalized nonlinear Schr\"{o}dinger equation~\cite{Akh97,Kiv03}
\begin{equation}
  {i \partial_z \psi} + {\beta \over 2 }\, \partial_x^2 \psi +
  (\gamma + i \gamma_a) |\psi|^2 \psi + i \alpha \psi = 0\, ,
\label{nlse}
\end{equation}
where $\psi(x, z)$ is the envelope of the electric field, $x$ and $z$
are the transverse and longitudinal coordinates, respectively, $\beta$
is the diffraction (dispersion) coefficient, $\gamma$ is the Kerr
nonlinearity parameter, $\gamma_a > 0$ characterizes nonlinear
absorption, and $\alpha < 0$ is the parameter of  linear gain.
Equation~(\ref{nlse}) can be normalized such that $\beta = \gamma = 1$,
therefore we use these values in all numerical simulations.

\begin{figure}[htbp]   
  \includegraphics[width=8cm]{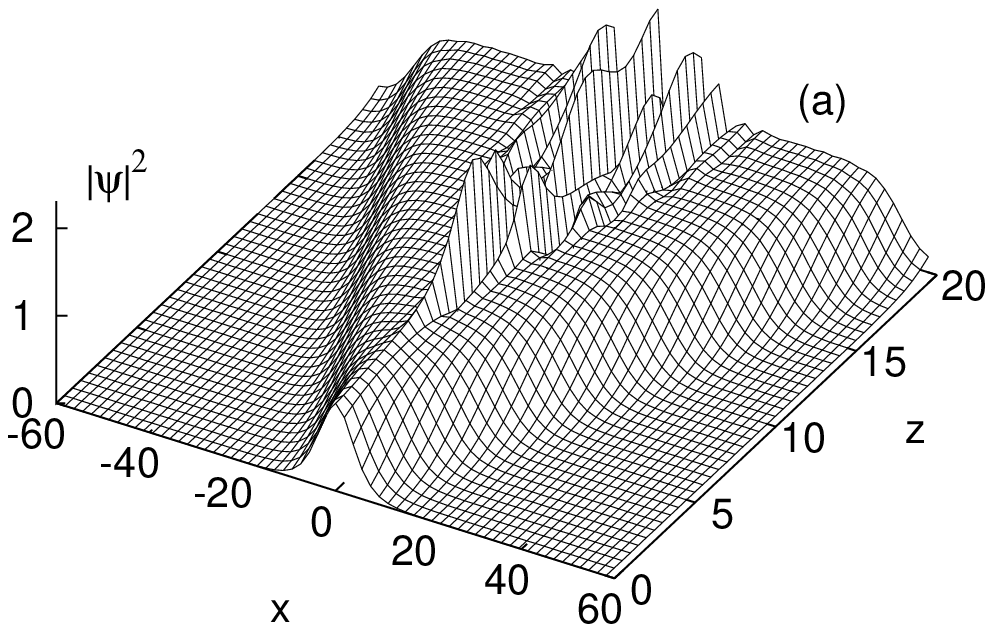}
  \includegraphics[width=8cm]{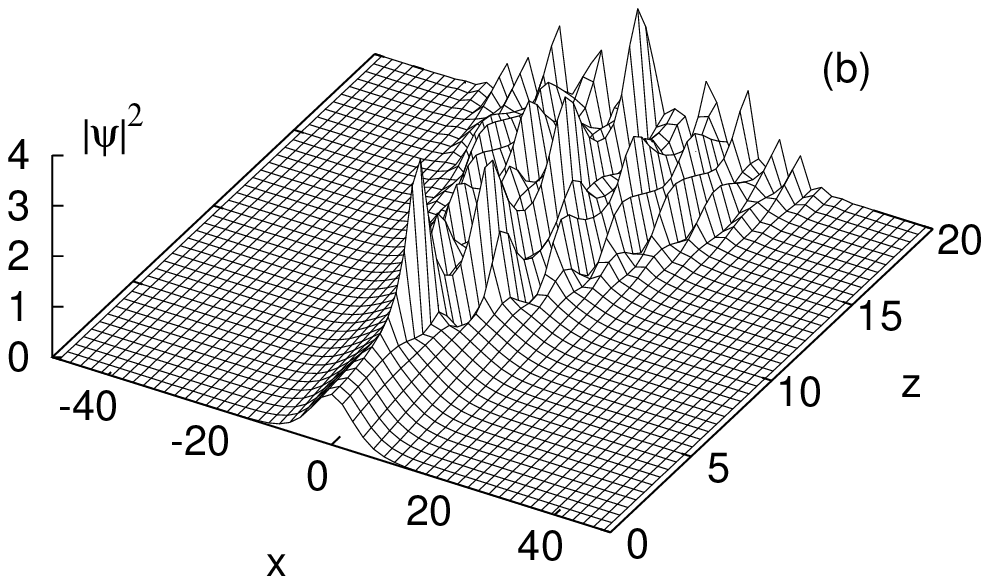}
\caption{The beam dynamics described by Eq.~(\ref{nlse}) for initial
condition $\psi(x,0)= \mbox{sech}(x/w_0)$. Plots show the
evolution of the beam intensity $|\psi|^2$. The parameters are (a)
$\alpha = -0.5$, $\gamma_a= 0.5$ and $w_0= 5$, (b) $\alpha = -0.2$,
$\gamma_a= 0.2$ and $w_0= 5$. } \label{fig:bd}
\end{figure}

  One can distinguish two basic scenarios of the beam propagation in
media with gain and losses, see Fig.~\ref{fig:bd}. Let us consider, for
example, a case of low initial peak intensity. Then, the peak intensity
of the beam increases initially due to linear gain. This increase is
limited by nonlinear dissipation. Intensity near the beam center keeps
at a constant value due to the balance between the two effects, while
intensity at the beam edges continues to rise. This process results in
a formation of two fronts moving in opposite directions, see
Fig.~\ref{fig:bd}(a). One can see that at large $z$, fronts move with
constant velocities as stationary waves.

  In the second scenario, Fig.~\ref{fig:bd}(b), the initial beam breaks
up into several pulses that in turn generate new pulses and so on.
Then, a lattice of pulses is created. That non-uniform region expands
into the unstable uniform one. In fact, a lattice of pulses is formed
in the first scenario as well, see Fig.~\ref{fig:bd}(a). The first
(second) scenario is realized when the front velocity is larger
(smaller) than that of the expansion of the lattice of pulses. In this
paper we are focused mainly on the first scenario, where the asymptotic
dynamics of the beam is determined by the front parameters. We provide
also a condition that separates the two types of the dynamics, see
Sec.~\ref{sec:ef}.

  The paper is organized as follows. The dynamical system for
stationary waves is  analyzed in Sec.~\ref{sec:sw}. In particular, the
equations for the front parameters are obtained there.
Section~\ref{sec:ef} discusses the condition of the front existence and
comparison with numerical simulations. The results are summarized in
Sec.~\ref{sec:c}.

\section{Stationary waves: Fronts}
\label{sec:sw}

  Fronts in Fig.~\ref{fig:bd} propagate with constant velocity,
therefore in this section we study stationary wave solutions of
Eq.~(\ref{nlse}). We follow the standard analysis described, for
example, in Ref.~\cite{Saa92}.  We look for solutions in the following
form:
\begin{equation}
  \psi(x,z)= a(\xi) \exp[i \phi(\xi)- i \mu z)],\quad \xi \equiv x - v z,
\label{form}
\end{equation}
where $v$ is the wave velocity, and $\mu$ is the propagation constant.
Substitution of Eq.~(\ref{form}) into Eq.~(\ref{nlse}) results in the
dynamical system for the wave parameters:
\begin{eqnarray}
  a' &=& p a,
\nonumber \\
  p' &=& q^2 - p^2 - {2 \over \beta}
  \left( v q + \mu + \gamma a^2 \right),
\nonumber \\
  q' &=& -2 p q + {2 \over \beta} \left( v p - \alpha -
   \gamma_a a^2 \right),
\label{ds}
\end{eqnarray}
where prime means $d/d \xi$, $p \equiv a' / a$, and $q= \phi\,'$. There
are two types of fixed points of Eqs.~(\ref{ds}), namely, those with
vanishing amplitude (``linear'' points), and those with finite
amplitude (``nonlinear'' points) (cf.~\cite{Saa92}):
\begin{eqnarray}
  a_{L} &=& 0, \quad p_L= \alpha / (v - \beta\, q_L) ,
\nonumber \\
  q_L &=& {1 \over \beta} \left[ v \pm
  \sqrt{ b +
  \sqrt{b^2 +
  \alpha^2 }}\, \right] ,
\label{fpl}\\
  b&\equiv& v^2/2 + \mu \beta,
\nonumber
\end{eqnarray}
and
\begin{eqnarray}
  a_N &=& \sqrt{-\alpha /\gamma_a}, \quad p_N= 0,
\nonumber \\
  q_N &=& {1 \over \beta} \left[ v \pm \sqrt{ v^2 +
  2 \beta \left(\mu  - \alpha \gamma / \gamma_a
  \right)}\; \right]
\label{fpn}
\end{eqnarray}
Without loss of generality, only points with non-negative amplitude $a
\ge 0$ are considered. The L-points each have one real eigenvalue and a
pair of complex conjugate eigenvalues:
\begin{equation}
  \lambda_1^{(L)} = p_L, \quad
  \lambda_{2,3}^{(L)} = -2 p_L \pm 2 i (q_L -v/\beta).
\label{ev1}
\end{equation}
The eigenvalues of each N-point are determined from the following
equation
\begin{equation}
  \lambda^3 + 4 \left( c_N^2 -
  {\alpha \gamma \over \beta \gamma_a} \right) \lambda -
  {8 \alpha \over \beta}\, c_N = 0, \quad c_N= q_N - v/\beta.
\label{cubic}
\end{equation}
If $c_N > 0$ ($c_N < 0$) then Eq.~(\ref{cubic}) has one negative
(positive) root and a pair of complex conjugate roots with the positive
(negative) real part.

  A localized wave of Eq.~(\ref{nlse}) corresponds to a separatrix of
dynamical system~(\ref{ds}). In particular, a pulse is described by a
separatrix that connects two L-points, while a front is described by a
separatrix that connects the L- and N-points.

  Equation~(\ref{form}) for stationary waves involves two unknown
parameters, $\mu$ and $v$. In general, there are no additional
conditions that fix these parameters. The theory of MS is developed for
dissipative systems with both dispersion and diffusion, i.e. when
parameter $\beta$ is complex. For purely dispersive media, when $\beta$
in Eq.~(\ref{nlse}) is real, the theory of MS gives that any value of
$v$ is possible.

  For Eq.~(\ref{nlse}), we find a result which is not typical for
dissipative systems. From our analysis, we conclude that the front
velocity is defined also by initial conditions. This is in contrast to
dissipative diffusive media, where $v$ depends mainly on the system
parameters.

  There are no regular methods to obtain the front parameters
for arbitrary initial conditions. Let us consider a particular initial
condition $\psi(x,0) \equiv \psi_0(x)= a_0(x) \exp [i \phi_0(x)]$ with
the following asymptotic behavior for amplitude $a_0(x)$ and phase
$\phi_0(x)$:
\begin{equation}
  a_0(x) \sim e^{- |x|/w_0},\ \ \phi'_0(x) \sim q_0
  \mbox{ at } |x| \to \infty.
\label{asym}
\end{equation}
We find that a front developed from such initial conditions moves
uniformly (see Fig.~\ref{fig:bd}(a)), while $v$ and $\mu$ are defined
from Eq.~(\ref{fpl}) with substitution
\begin{equation}
  p_L= \mp 1/w_0, \quad q_L= q_0,
\label{subs}
\end{equation}
or explicitly
\begin{eqnarray}
  v &=& \mp \alpha\, w_0 + \beta q_0,
\nonumber\\
  \mu &=& \pm \alpha\, w_0 q_0 - {\beta \over 2} \left(q_0^2 + 1/w_0^2
  \right).
\label{vmu}
\end{eqnarray}
If $a= 0$ is at $x = +\infty$ ($x = -\infty$) then the upper (lower)
sign in Eqs.~(\ref{subs}) and~(\ref{vmu}) should be chosen. Therefore,
we come to an interesting result that the parameters of a stationary
front are governed by the parameters of the L-point only, or, in other
words, by the parameters of small-amplitude waves of Eq.~(\ref{nlse}).

  In the equation for $v$, the second term is the phase velocity
of linear waves. Then, the first term is the relative velocity of the
front in the corresponding reference frame. We use this relative
velocity in Sec.~\ref{sec:ef}.

\begin{figure}[htbp]   
  \includegraphics[width=7.5cm]{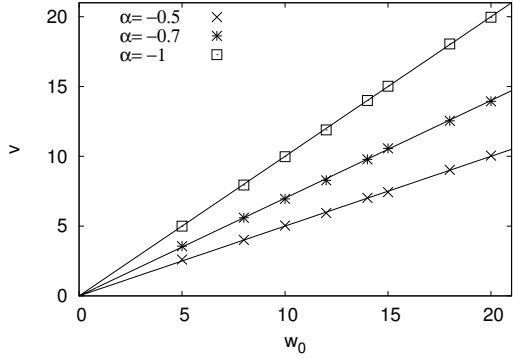}
\caption{The dependence of the front velocity on $w_0$ for $\gamma_a=
0.5$ and different values of $\alpha$. Points correspond to result of
numerical simulations of Eq.~(\ref{nlse}). Lines are found from
Eqs.~(\ref{vmu}). } \label{fig:fv}
\end{figure}

  Figure~\ref{fig:fv} shows good agreement between the values of the
front velocity found from numerical simulation of Eq.~(\ref{nlse}) and
calculated using Eqs.~(\ref{vmu}). The initial condition is taken as
\begin{equation}
  \psi_0(x)= a_N\, \mbox{sech}(x/w_0).
\nonumber
\end{equation}
The velocity is determined numerically for large $z$, when all
transient processes are ended. All points in Fig.~\ref{fig:fv}
corresponds to the first scenario, when well-pronounced fronts are
developed, see Fig.~\ref{fig:bd}(a). This scenario is realized for
sufficiently large $w_0$. This is discussed in detail in
Sec.~\ref{sec:ef}.

\begin{figure}[htbp]   
  \includegraphics[width=8cm]{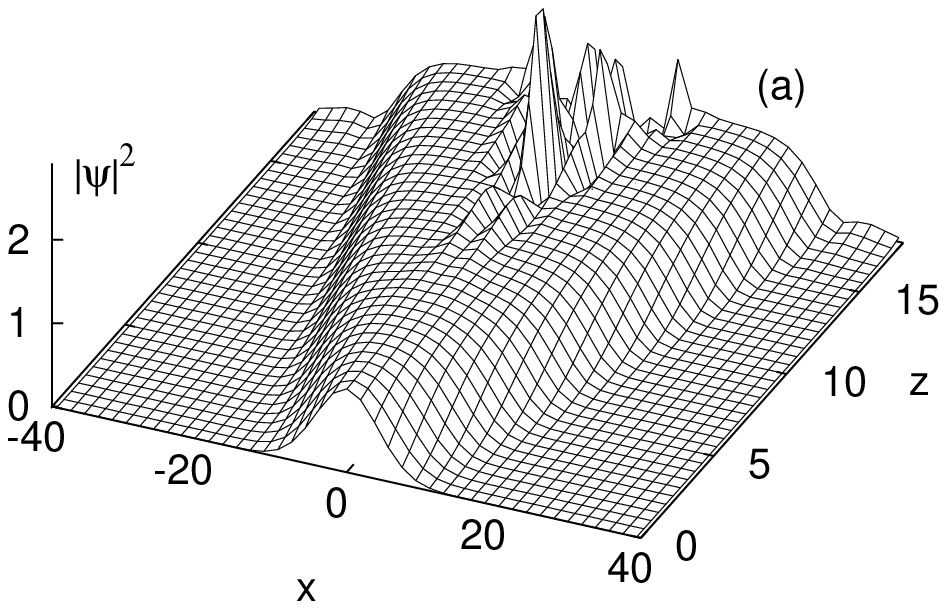}
  \includegraphics[width=8cm]{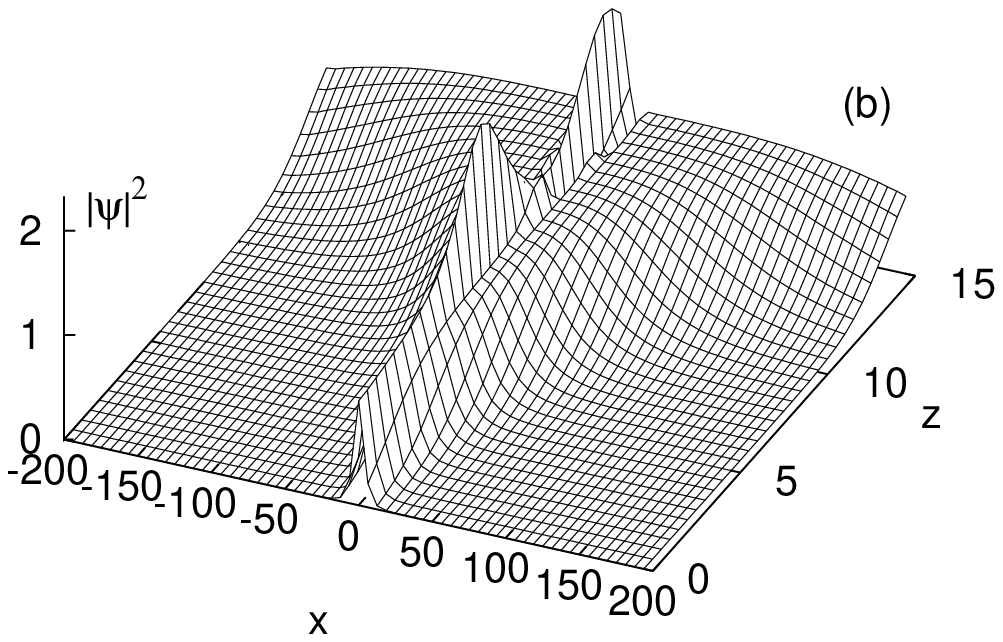}
\caption{The dynamics of (a) a Gaussian pulse, $\psi(x,0)=
\exp[-(x/6)^2/2]$, and (b) a Lorentzian pulse, $\psi(x,0)=
[(x/6)^2+1]^{-1}$. The parameters are $\alpha =-0.5$ and $\gamma_a=
0.5$. } \label{fig:gl}
\end{figure}

  We extend our results further to beams with asymptotic behavior
different from that in Eq.~(\ref{asym}). Let us consider for simplicity
the real $\psi_0(x)$. If $\psi_0(x)$ decreases at $|x| \to \infty$
slower (faster) than exponentially, then the emerging fronts move
decelerating (accelerating).  Figure~\ref{fig:gl} represents such
dynamics for a Gaussian pulse and a Lorentzian pulse, respectively. We
obtain that even in these non-stationary cases, Eqs.~(\ref{fpl})
and~(\ref{subs}) can still be used for estimation of front parameters,
provided that $w_0$ and $q_0$ are corresponding function of $z$.

  The analysis above shows that media with linear amplification and
nonlinear dissipation can be used to distinguish beams with different
asymptotics. A deviation of the beam's asymptotic behavior from an
exponential one results in a deviation of the dependence of the beam
width on $z$ from a straight line.

\section{Existence of fronts}
\label{sec:ef}

  There are two questions regarding the front existence. The first
question is for what values of the system parameters does the L-N
separatrix exist. The second question is what are the conditions for
development of well-pronounced fronts. We discuss these question in the
present section.

  There are no general methods to determine the existence of L-N
separatrices. However, our study uncover a notable property of
system~(\ref{ds}). It is known~\cite{Noz84} that Eq.~(\ref{nlse}) has
an exact soliton solution. It corresponds to a L-L separatrix of
Eqs.~(\ref{ds}). On the other hand, numerical simulations of
Eq.~(\ref{nlse}) show unambiguously the existence of fronts, or L-N
separatrices, for other sets of the system parameters. Also, analysis
of the eigenvalues of the L-points reveals that only a single
trajectory goes out of plane $a= 0$ from the L-point. It means that the
L-L and L-N separatrices cannot co-exist for a given set of the
parameters. These facts indicate a presence of a {\em global}
bifurcation~\cite{Guc83} in system~(\ref{ds}), and therefore in
Eq.~(\ref{nlse}). Namely, depending on the parameters, the separatrix
that starts from the L-point, ends at either the another L-point or the
N-point as $\xi \to \pm \infty$, depending on the parameters. At the
same time, the local properties of model~(\ref{ds}), in particular, the
types of the fixed points, do not change near the bifurcation.

\begin{figure}[htbp]   
  \includegraphics[width=9cm]{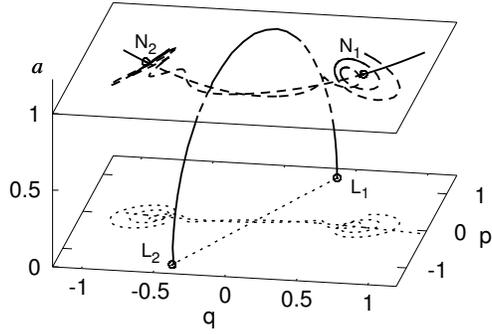}
  \includegraphics[width=9cm]{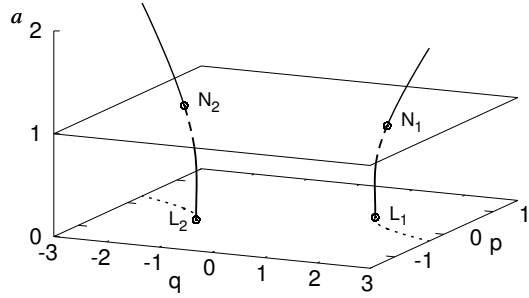}
\caption{Separatrices in the phase space of Eqs.~(\ref{ds}) for  $v=
0$, $\alpha =-0.5$, and $\gamma_a= 0.5$. The L-points (N-points) are
labelled as $L_1$ and $L_2$ ($N_1$ and $N_2$). (a) The L-points are
connected via the L-L separatrix at $\mu= -0.7104$. (b) There are two
L-N separatrices at $\mu= 0.85$. The dotted lines show the projections
of the separatrices on plane $a= 0$. } \label{fig:ps}
\end{figure}

  Let us fix all the parameters in Eqs.~(\ref{ds}), except of $\mu$. We
find that the L-L (L-N) separatrix exists below (above) a bifurcation
value $\mu_B$. Figure~\ref{fig:ps} shows separatrices that go from (to)
the L- and N-points for different sets of the parameters. Only
trajectories with real eigenvalues, $\lambda_L$ and $\lambda_N$, are
shown. In order to reconstruct the heteroclinic trajectories in
Fig.~\ref{fig:ps}, we take an initial condition at the vicinity of the
fixed point along the direction of the corresponding eigenvector. Each
trajectory is checked by integrating backward on $\xi$ from the end
point.

  In Fig.~\ref{fig:ps}(a), $\mu= -0.7104 < \mu_B$, there is a
separatrix that connects the two L-points. One separatrix of the
N-point is located in a region $a > 1$. The other separatrix of the
N-point  goes to (from) a region $a < 1$, intertwining with the
separatrix of the second N-point. In Fig.~\ref{fig:ps}(b), $\mu= 0.85 >
\mu_B$, there is a separatrix that connect points of distinct types.

  The value $\mu= -0.7104$ for $v= 0$, $\alpha= -0.5$ and $\gamma_a =0.5$
corresponds to the exact soliton solution~\cite{Noz84}. The value $\mu=
0.85\equiv \overline{\mu}_B(v=0)$ for the same values of the other
parameters can be considered as an upper estimate of $\mu_B$. We find
$\overline{\mu}_B(0)$ such that integrating forward or backward on
$\xi$ from the one of the fixed point, the trajectory approaches the
other fixed point closer than 0.002 in absolute units. For other values
of $v$, the bifurcation value $\mu_B$ (or $\overline{\mu}_B$) can be
calculated using the Galilean invariance, $\mu_B(v)= \mu_B(0)- v^2/2$.
We also find that $\overline{\mu}_B$ depends weakly on $\gamma_a$ and
almost linearly on $\alpha$.

  Now we turn to the question of the front development. First, we
analyze {real} initial conditions, i.e. without phase modulations. We
find that the first scenario with well-defined fronts is realized, when
the front velocity is larger than the threshold velocity $v_{th}$. The
threshold $v_{th}$ is the velocity of an expansion of the pulse region.
The later, we believe, is related to the modulational instability of
the plane wave with amplitude $a_N$. Namely, a tail of a pulse plays a
role of a perturbation of the uniform (L- or N-) state. Due to
instability,  new pulses are generated from this perturbation resulting
in the expansion of the pulse region.

  In order to estimate $v_{th}$, we perform the analysis of the
modulational instability of Eq.~(\ref{nlse}).  Let us consider the
plane-wave solution of Eq.~(\ref{nlse}):
\begin{equation}
  \psi(x, z)= a_N e^{i\varphi(x)}, \quad \varphi(x)= (k x - \omega z),
\label{pw}
\end{equation}
where
\begin{equation}
  \omega= \beta k^2 /2 + \alpha \gamma /\gamma_a.
\nonumber
\end{equation}
Equation for small modulations $u(x, z)$ can be obtained by
substitution of the field in the form
\begin{equation}
  \psi= [a_N + u(x,z)]e^{i\varphi(x)}
\nonumber
\end{equation}
into Eq.~(\ref{nlse}) and linearizing on $u$. Then, assuming $u \sim
\exp[i(K x - \Omega z)]$, one can derive the dispersion relation of
modulations:
\begin{equation}
  \Omega= \beta k K + i \alpha \pm
  \left[ - \alpha^2 + {\beta K^2 \over 2}
  \left( {\beta K^2 \over 2} + 2 {\alpha \gamma \over \gamma_a}
  \right) \right]^{1/2}.
\label{dr}
\end{equation}
It follows from the analysis of Eq.~(\ref{dr}) that the plane wave
(\ref{pw}) is unstable when $|K| < 2 [-\alpha \gamma/(\beta
\gamma_a)]^{1/2}$. The instability gain, $g(K)\equiv
\mbox{Im}[\Omega(K)]$, has a maximum at $K= K_m= [-2 \alpha \gamma
/(\beta \gamma_a)]^{1/2}$. Then the characteristic velocity $v_{MI}$ of
the instability expansion is estimated as a ratio of characteristic
scales on $z$ and $x$, namely the maximum of instability gain and the
corresponding wavenumber:
\begin{equation}
  v_{MI} \equiv {g(K_m) \over K_m}.
\end{equation}

  Numerical simulations of Eq.~(\ref{nlse}) reveal that the threshold
velocity can be determined as  $v_{th}= C v_{MI}$. Fitting of this
equation with numerical data gives $C \approx 4.7$ for $\beta =\gamma =
1$. Then, well-defined fronts exist if
\begin{equation}
  |\alpha w_0|  > v_{th} \equiv
  C \sqrt{|\alpha| \beta \gamma_a \over
  2 \gamma} \left( \sqrt{1 + \gamma^2/\gamma_a^2} - 1 \right).
\label{vth}
\end{equation}
This condition defines the parameter $w_0$ of the initial profile that
develops into two fronts.

\begin{figure}[htbp]   
  \includegraphics[width=7.5cm]{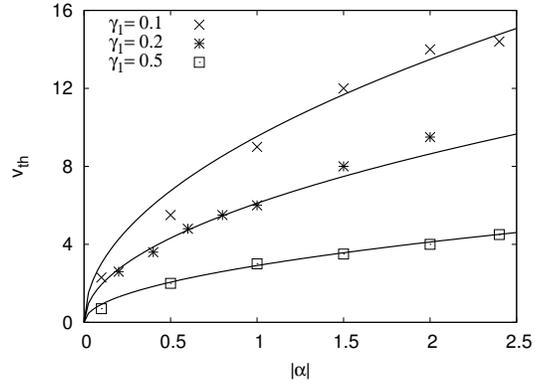}
\caption{The threshold velocity as a function of $|\alpha|$ for
different $\gamma_a$. Points correspond to numerical simulations of
Eq.~(\ref{nlse}). Lines correspond to Eq.~(\ref{vth}), $C\approx 4.7$.
} \label{fig:vt}
\end{figure}

  The dependence of $v_{th}$ on the system parameters found numerically
and using Eq.~(\ref{vth}) is shown in Fig.~\ref{fig:vt}. For a given
set of the system parameters, we change the initial width $w_0$, until
the velocity of the emerging fronts exceeds the velocity of pulse
spreading. The threshold is not sharp, therefore the relative error of
$v_{th}$ is in order 0.1. Nevertheless, we see a good agreement between
numerical results and Eq.~(\ref{vth}). We also calculate numerically
the propagation constant $\mu_{th}$ of the front at $v= v_{th}$. Taking
$\mu$ as a control parameter, we find that the formation of
well-defined fronts occurs at larger $\mu$ than the global bifurcation,
$\mu_{th} > \overline{\mu}_B(v_{th})$.

  An addition of the linear phase to initial conditions results
in a constant shift of the whole beam profile with the corresponding
phase velocity. This is a manifestation of the Galilean invariance
discussed in Sec.~\ref{sec:sw}. In this case one should compare the
relative front velocity and $v_{th}$, obtaining the same
Eq.~(\ref{vth}). Therefore, Eq.~(\ref{vth}) is valid also for initial
conditions with linear modulation of the phase.

\section{Conclusion}
\label{sec:c}

  We have demonstrated that in dispersive media with gain and losses,
a beam propagates for some set of the system parameters in a form of
two fronts moving in opposite directions. It has been shown that the
front parameters depend on the system parameters, as well as on initial
conditions. The presence of global bifurcation of stationary waves in
Eqs.~(\ref{ds}) is found. The threshold for development of fronts has
been obtained. A possibility to use dispersive media to distinguish
beams with different asymptotic behavior has been suggested.




\newpage

\end{document}